\documentclass{article}

\usepackage{fullpage}
\usepackage{authblk}

\usepackage{tikz}

\usepackage{graphicx,amsmath,amssymb,amsfonts,amsthm,mathtools} 
\usepackage[dvipsnames]{xcolor}
\usepackage[numbers,sort]{natbib}
\usepackage{csquotes}
\usepackage{booktabs}
\usepackage{enumerate}

\definecolor{linkcolor}{rgb}{0,0.2,0.6}
\usepackage{hyperref}
\hypersetup{colorlinks,breaklinks,urlcolor=linkcolor,linkcolor=linkcolor,citecolor=linkcolor}

\newtheorem{theorem}{Theorem}[section]

\theoremstyle{definition}
\newtheorem{definition}{Definition}

\theoremstyle{remark}
\newtheorem*{remark}{Remark}
\newtheorem*{example}{Example}

\title{Phylogeny-based metrics of biodiversity: concepts and methods}
\author[1,2]{Kristina Wicke \thanks{\url{kristina.wicke@njit.edu}}}
\author[3]{Arne Mooers \thanks{\url{amooers@sfu.ca}}}

\affil[1]{Department of Mathematical Sciences, New Jersey Institute of Technology, Newark, NJ, USA}
\affil[2]{National Institute for Theory and Mathematics in Biology, Northwestern University and The University of Chicago, Chicago, IL 60611, USA}
\affil[3]{Department of Biological Sciences, Simon Fraser University, Vancouver, BC V5A 1S6, Canada}

\begin{document}

\date{}
\maketitle

\begin{abstract}
The concept of valuing evolutionary history has a long tradition and was formalized by several research groups in the early 1990s, each aiming to capture the products of evolution more comprehensively than species richness alone. Preserving the \enquote{Tree of Life} has since become a central, if sometimes contested, goal in conservation biology. Faith’s phylogenetic diversity (PD), defined as the sum of the edge lengths of a rooted tree connecting a set of focal entities, is the most widely used metric in this context and underpins an extensive methodological framework. This framework includes widely used species-specific indices such as EDGE, as well as approaches that quantify expected species-specific contributions to PD such as EDGE2.

This chapter focuses on the development of this framework of phylogeny-based conservation metrics. We review the development of species-specific indices, highlighting their strengths and limitations, and trace their progression from simple formulations to game-theoretic approaches and more abstract generalizations. We also discuss how these indices are applied in conservation practice, with particular emphasis on the EDGE framework. We further discuss the relationship between the edge lengths of a phylogeny and the evolutionary features they are intended to represent. Finally, we highlight recent advances in the field and identify areas for future research.
\end{abstract}


\section{Introduction} \label{sec:introduction}
The international conservation community has valued evolutionary history since at least 1980, when, in a landmark volume, The International Union for the Conservation of Nature (IUCN) suggested, inter alia, that species in monotypic genera and monotypic families should receive priority attention over those found in polytypic taxa by virtue of their distinctiveness (\citet{IUCN1980}).  This taxonomic approach to prioritization was made explicit policy supporting the  Endangered Species Act in 1983 (\citet{FWS1983}), and the general argument was made clear by \citet{Atkinson1989}: \enquote{Given two threatened taxa, one a species not closely related to other living species and the other a subspecies of an otherwise widespread and common species, it seems reasonable to give priority to the taxonomically distinct form.} The argument combines the key concept of evolutionary redundancy with a clear goal to maximize total evolutionary history via targeted conservation.

The logic behind this goal was made somewhat explicit in a landmark paper by \citet{Faith1992}: because of divergent evolution and heritability, the branching pattern of an underlying phylogeny predicts the distribution of character states at the tips, such that a set of tips that spans more of the encompassing phylogeny (i.e., is less redundant) will exhibit a broader range of such states (see Figure~\ref{fig:tree}). This broader range is assumed to offer more \enquote{option value} to humans. Faith coined the term \enquote{phylogenetic diversity} or \enquote{PD} to capture the total path length of the tree connecting a set of $k$ tips to some specified root, and this total path length is evolutionary history. Importantly, while the tips on a phylogenetic tree are most often species, they could be other entities, e.g., populations. For simplicity, we will use \enquote{species} to refer to any entity represented by a leaf on a tree or network.

Two additions are required to tie this framework back to taxonomic prioritization. The first is a species' contribution to PD. \citet{Altschul1990} noted that one could use the length of the pendant edge on a complete phylogeny as a metric of \enquote{degree of change} and so worth, and this was echoed by \citet{Faith1992} in the context of non-redundancy.  Atkinson's implied argument quoted above is that a species not closely related to others sits on a longer pendant edge than does a subspecies, and so contributes more to evolutionary history and so to the total range in character states.    

The second important addition is that of \enquote{expected PD loss,} or the amount of present-day evolutionary history projected to be lost to extinction at some point in the future. This concept, presented clearly in a two-page note by \citet{Witting1995} (see also \citet{Witting1994}), is based on the fact that an internal edge in a phylogeny is only lost if all the species it gives rise to are lost.  Therefore, the expected loss of an edge in a phylogenetic tree is simply its length (its weight) multiplied by the product of the extinction probabilities of all the species that subtend it; the sum of these terms across all edges in a tree is the expected (or average) loss of phylogenetic diversity. In turn, one can adjust every edge length by subtracting its expected loss contribution to construct an \enquote{expected PD} tree. With this framework, \citet{Witting1995} describe how to choose subsets (e.g., choose among protected areas that house different species) to minimize the expected loss of evolutionary  history, i.e., maximize the expected PD tree.

In a pair of papers published in 1992 and 1993 (\citet{Weitzman1992,Weitzman1993}), the late economist Martin Weitzman independently derived a PD framework for measuring conservation-relevant diversity and then used it to measure the expected contributions of individual species following extinction. Weitzman's metric, though couched in the economics terminology of \enquote{elasticities} and \enquote{expected present discounted diversity} maps directly onto the EDGE2 metric we review in a later section.

Taken together, these developments established the foundations of phylogeny-based conservation while also giving rise to a diverse and expanding set of metrics for quantifying evolutionary history and its conservation value. The aim of this chapter is to review this body of work, tracing the development of phylogeny-based conservation metrics from simple indices to abstract formulations.
After introducing phylogenetic trees (Section~\ref{sec:trees}) and briefly revisiting Daniel P. Faith’s original definition of phylogenetic diversity (PD) (Section~\ref{sec:PD}), we discuss a range of species-specific PD indices in Section~\ref{sec:indices}. These range from relatively simple measures, such as the equal splits and fair proportion indices originally used in the EDGE framework, to more complex indices that incorporate the expected PD framework, particularly the EDGE2 index. We then review prominent abundance-weighted measures of biodiversity (Section~\ref{sec:abundance}), including Rao's quadratic entropy and phylogenetic Hill numbers. Section~\ref{sec:networks} is devoted to recent extensions of PD and related indices to structures more general than phylogenetic trees, namely phylogenetic networks. Because PD and its associated indices depend on the edge lengths of a phylogeny, whether tree or network, Section~\ref{sec:lengths} discusses what these edge lengths represent and the implications for interpreting PD. We conclude with a brief discussion highlighting future directions for the field and areas that warrant further investigation.

\section{Phylogenetic trees} \label{sec:trees}

\subsection{Trees as representations of evolutionary history}
\begin{figure}[htbp]
    \centering
    \includegraphics[width=0.5\linewidth]{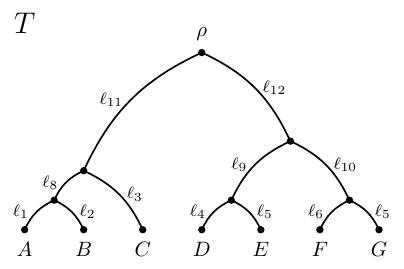}
    \caption{A rooted binary phylogenetic $X$-tree with $X = \{A, \ldots, G\}$. Leaves represent observed species, internal nodes correspond to hypothetical common ancestors, and edge lengths indicate the amount of evolutionary change or time separating lineages.}
    \label{fig:tree}
\end{figure}

Several measures of phylogenetic diversity are based on a graphical representation of evolutionary relationships among species, most commonly in the form of a rooted phylogenetic tree. Intuitively, such a tree represents evolutionary history, where lineage-splitting events correspond to branching points (internal nodes) and descent with modification is represented along the edges connecting them. This representation provides a way to encode shared ancestry among species and forms the basis for quantifying the amount of evolutionary history contained within a set of taxa. An example of such a tree is shown in Figure~\ref{fig:tree}. In this example, the leaves correspond to observed species (labeled $A$-$G$), while internal nodes represent hypothetical common ancestors. The root $\rho$ represents the most recent common ancestor of all taxa shown.

\subsection{Formal definitions and notation}
Let $X$ denote a non-empty finite set of taxa. A \emph{rooted phylogenetic $X$-tree} is a tree $T = (V(T), E(T))$ with a distinguished root $\rho$, in which the leaves are bijectively labeled by the elements of $X$, and each internal node represents a divergence event. Unless otherwise specified, we assume that each internal node has at least two descendants; if every internal node has exactly two descendants, the tree is said to be \emph{binary}.

Each edge $e \in E(T)$ is typically assigned a non-negative length $\ell(e)$, which plays a central role in defining measures of phylogenetic diversity.
These lengths can be interpreted in one of several ways, for example as representing evolutionary time, the amount of genetic change, or more abstractly the accumulation of evolutionary features along a lineage. 

A phylogenetic tree is called \emph{ultrametric} if all leaves are equidistant from the root. Ultrametric trees typically arise under molecular clock assumptions, where edge lengths represent evolutionary time. Non-ultrametric trees, by contrast, allow leaves to have different distances from the root, often reflecting variation in the rates of molecular or phenotypic change used to infer the trees. This distinction is important for interpreting measures of phylogenetic diversity: in ultrametric trees, longer edges reflect deeper evolutionary history in time, whereas in non-ultrametric trees, longer edges may reflect greater accumulated change or feature diversity if the edge lengths predict these quantities more generally. We return to the role and interpretation of edge lengths in more detail in Section~\ref{sec:lengths}.

\subsection{Inference of phylogenetic trees and uncertainty}
Phylogenetic trees are not directly observed but are inferred from data, typically molecular sequence data or, more rarely, morphological traits. A variety of statistical and algorithmic methods are used for this purpose, including distance-based approaches, maximum likelihood estimation, and Bayesian inference (see, e.g., \citet{Felsenstein2004,Lemey2009,Stadler2024} for further details). These methods differ in their assumptions and in how they quantify uncertainty, but all produce estimates of both tree topology and edge lengths. As a result, phylogenetic trees should be viewed as inferred objects, subject to model assumptions and estimation error. This uncertainty can propagate into downstream diversity measures, although in practice it is often ignored.

\section{Faith's phylogenetic diversity (PD)} \label{sec:PD}
In the early 1990s, Daniel P. Faith published a seminal paper demonstrating how phylogenetic trees can be used to identify sets of species or populations that maximize what he termed ``feature diversity" (\citet{Faith1992}). His primary motivation was to inform biodiversity conservation decisions under limited resources, following the assumption that preserving a broad range of features is a desirable objective. Building on earlier work by \citet{Vanewright1991}, which discussed the importance of taxonomic distinctiveness in conservation prioritization, Faith introduced the concept of \emph{phylogenetic diversity} (PD).

Phylogenetic diversity is defined as the sum of edge lengths in the minimal subtree connecting a chosen set of leaves to the root of a given phylogenetic tree. 

\begin{example}
    Consider the tree $T$ in Fig.~\ref{fig:tree}. For the subset $Y = \{A,B,C\}$, the phylogenetic diversity is $PD_T(Y) = \ell_1 + \ell_2 + \ell_3 + \ell_8 + \ell_{11}$.
    By contrast, for $Y' = \{A,C,D\}$, we have $PD_T(Y')= \ell_1 + \ell_3 + \ell_4 + \ell_8 + \ell_9 + \ell_{11} + \ell_{12}$.
\end{example}

In Faith's original formulation, edge lengths were measured as the number of inferred character changes under a maximum parsimony reconstruction on a cladogram derived from a character-state matrix without homoplasy\footnote{Homoplasy is a biological phenomenon in which different species share similar traits that did not arise from a common ancestor, but instead evolved independently (e.g., the wings of birds and bats).}. Faith showed, through an illustrative example, that summing these reconstructed edge lengths yields the same total feature diversity as directly counting features from the underlying character matrix. 

Although it is now well understood that PD coincides with observed feature diversity only under specific conditions (\citet{Wicke2021}), it has nonetheless become one of the most widely adopted measures of biodiversity in both theoretical and applied settings, with Faith's original paper cited almost 7,000 times.

This popularity likely stems in part from the simplicity and interpretability of PD: under the assumption that the edge lengths are predictive of feature evolution, maximizing PD can be viewed as preserving as much of the \enquote{Tree of Life} as possible. Beyond this intuitive appeal, PD also enjoys a number of useful mathematical properties, including \emph{monotonicity} and \emph{submodularity}. 
Monotonicity captures the idea that adding species cannot decrease diversity. Formally, if $T$ is a phylogenetic $X$-tree and $Y' \subseteq Y \subseteq X$, then $PD_T(Y') \leq PD_T(Y)$. 
Submodularity, by contrast, is the property that for all $Y, Y' \subseteq X$, 
\[ PD_T(Y \cup Y') + PD_T(Y \cap Y') \leq PD_T(Y) + PD_T(Y').\]

From an optimization perspective, the problem of selecting a subset of $k$ species that maximizes PD admits an efficient solution via a simple greedy algorithm, which iteratively adds the species with the largest marginal contribution to PD (\citet{Steel2005,Pardi2005}). However, this computational tractability is lost under more realistic modeling assumptions. In practice, conservation decision often involve heterogeneous conservation costs across species together with a fixed budget, leading to a setting in which maximizing PD becomes NP-hard (\citet{Moulton2007}). Similarly, incorporating additional ecological structure can further increase computational complexity. For example, accounting for dependencies among species through food webs or other interaction networks introduces constraints that likewise render the problem NP-hard (see, e.g., \citet{Moulton2007,Faller2011,Pardi2007}).

These challenges highlight a limitation of the subset selection framework, as optimal solutions can be difficult to compute and sensitive to modeling assumptions. This has motivated an alternative approach based on species-specific \emph{phylogenetic diversity indices}, which shift the focus from selecting subsets to ranking individual species based on their contribution to PD.

\section{Phylogenetic diversity indices} \label{sec:indices}
While phylogenetic diversity (PD) measures the total diversity represented by a set of taxa, PD indices instead assign each species a single numerical value reflecting its relative contribution to overall PD. One motivation for this approach is conservation planning, as it allows species to be ranked according to their contributions to total PD, providing a simple basis for prioritization decisions, even if the individual decisions do not lead to an optimal overall strategy for preserving PD.

There are multiple ways to measure relative contribution to total PD across extant species, and accordingly a range of PD indices have been developed for phylogenetic trees (for an overview, see, e.g.,~\citet{Vellend2011,Redding2014}).

\subsection{Simple PD indices}  \label{sec:simple}
Two particularly simple phylogenetic diversity (PD) indices that depend only on a rooted phylogenetic $X$-tree with edge lengths, and require no additional inputs such as extinction probabilities or conservation costs, are the \emph{equal splits} (ES) index and the \emph{fair proportion} (FP) index (also known as \emph{evolutionary distinctiveness} (ED) score). 

Both indices satisfy the \emph{efficiency property}: when summed over all species $x \in X$, their values equal the total edge length of the tree, i.e., 
\[ \sum_{x \in X} \text{Index}_T(x) = \sum\limits_{e \in E(T)} \ell(e) = PD_T(X).\]
Thus, each index provides a partition of total phylogenetic diversity across the leaves of the tree. 
Moreover, both indices are linear functions of the edge lengths.

In what follows, let $T$ be a rooted phylogenetic $X$-tree with root $\rho$, and let $P(T; \rho, x)$ denote the unique path in $T$ from $\rho$ to leaf $x \in X$. 

\paragraph*{Equal splits index.}
The equal splits index assigns to each leaf a share of the length of every edge on $P(T; \rho, x)$, dividing that length equally at each branching point along the path.

\begin{definition}[Equal splits index (\citet{Redding2003})]
    Let $T$ be a rooted phylogenetic $X$-tree. The \emph{equal splits index} of a leaf $x \in X$ is 
    \begin{align} \label{eq:es}
        ES_T(x) &= \sum\limits_{e \in P(T; \rho, x)} \frac{\ell(e)}{\Pi(e,x)},
    \end{align}
    where $\Pi(e,x)=1$ if $e$ is the pendant edge incident with $x$, and otherwise, for an interior edge $e=(u,v)$, $\Pi(e,x)$ is the product of the out-degrees of the interior vertices on the directed path from $v$ to $x$.
\end{definition}

\paragraph*{Fair proportion index.}
The fair proportion index also sums contributions from edges on  $P(T; \rho, x)$, but instead divides the length of each edge equally among all descendant leaves.

\begin{definition}[Fair proportion index (\citet{Redding2003,Isaac2007})]
    Let $T$ be a rooted phylogenetic $X$-tree. The \emph{fair proportion index} of a leaf $x \in X$ is  
    \begin{align} \label{eq:fp}
         FP_T(x) &= \sum\limits_{e \in P(T; \rho, x)} \frac{\ell(e)}{n(e)},
    \end{align}
    where $n(e)$ denotes the number of leaves descended from the edge $e$.
\end{definition}

\begin{example}
    Consider again the tree in Fig.~\ref{fig:tree}. For the leaf $A$, we have
    \[ES_T(A) = \ell_1 + \frac{\ell_8}{2} + \frac{\ell_{11}}{4}, \]
    whereas 
    \[ FP_T(A) = \ell_1 + \frac{\ell_8}{2} + \frac{\ell_{11}}{3}.\]
    Thus, the two indices differ in the contribution assigned to the edge of length $\ell_{11}$: the equal splits index divides this edge length according to successive branching events, while the fair proportion index allocates it equally among all descendant leaves.
\end{example}

\paragraph*{EDGE framework.}
Both the ES and FP indices induce a natural ranking of taxa that can be used to inform conservation decisions: species with higher index values are interpreted as contributing more to overall phylogenetic diversity, and this is likely generally true (\citet{Steel2018,Chaudhary2018}).  Although this approach may appear overly simplistic, it has been implemented in practice for the fair proportion index. In particular, the FP index was originally employed in the \enquote{EDGE of Existence} programme established by the Zoological Society of London, a conservation initiative that focuses on threatened species representing a high degree of unique evolutionary history. 

Specifically, the EDGE score of a species $x \in X$ is defined as
\[ EDGE(x) = \ln (1 + FP(x)) + GE(x) \cdot \ln(2),\]
where $GE(x)$ is denotes the IUCN Red List category weight assigned to $x$.

Following~\citet{Isaac2007}, species classified in threatened IUCN Red List categories, namely vulnerable (VU), endangered (EN), or critically endangered (CR), and having above-median FP values within their clade were identified as priority \enquote{EDGE species.} Particular conservation attention was then given to the highest-ranking species (e.g., the top 100, 50, or 25) within specific taxonomic groups. 

This approach has been used to generate EDGE lists for a wide range of taxa, including mammals, amphibians, birds, corals, reptiles, gymnosperms, and sharks and rays. More broadly, the EDGE framework has informed direct conservation actions for many of the identified species and the Zoological Society of London's EDGE of Existence programme has supported over 120 conservation projects worldwide targeting priority EDGE species (\citet{Gumbs2023}). In 2023, an updated protocol, EDGE2, was proposed to replace the original framework (\citet{Gumbs2023}). While it continues to rely on phylogenetic diversity indices, it explicitly incorporates uncertainty and extinction risk (see Section~\ref{sec:edge2}).

\paragraph*{Shapley value.}
Before turning to PD indices that incorporate extinction risk and the EDGE2 framework, we briefly introduce the Shapley value from cooperative game theory. The Shapley value provides a canonical method for distributing a total payoff among players based on their marginal contributions. In the context of phylogenetic diversity, this viewpoint leads to a curious result: the fair proportion index coincides with the Shapley value of the associated diversity game.

Formally, a \emph{cooperative game} is a pair $(Y,\nu)$, where $Y$ is a finite set of players and $\nu:2^Y \to \mathbb{R}$ is a characteristic function satisfying $\nu(\emptyset)=0$. A function $\varphi_\nu : Y \to \mathbb{R}$ assigning a payoff to each player is called a \emph{value} of the game. An important example is the \emph{Shapley value} (\citet{Shapley1953}), defined for each $i \in Y$ by
\[ \varphi_\nu(i) = \frac{1}{|Y|!} \sum\limits_{C \subseteq Y: i \in C} (|C|-1)! (|Y|-|C|)! (\nu(C) - \nu(C \setminus \{i\})).\]
Thus, $\varphi_\nu(i)$ represents the average marginal contribution of player $i$ over all possible orders in which the grand coalition $Y$ can be formed.

\noindent The Shapley value is uniquely characterized by the following four axioms:
\begin{enumerate}
\item \emph{Efficiency (Pareto efficiency):} $\sum_{i \in Y} \varphi_\nu(i) = \nu(Y)$.
\item \emph{Symmetry:} If $i,j \in Y$ are such that $\nu(C \cup \{i\}) = \nu(C \cup \{j\})$ for all $C \subseteq Y \setminus \{i,j\}$, then $\varphi_\nu(i) = \varphi_\nu(j)$.
\item \emph{Dummy player:} If $\nu(C \cup \{i\}) = \nu(C)$ for all $C \subseteq Y \setminus \{i\}$, then $\varphi_\nu(i)=0$.
\item \emph{Additivity:} For any two characteristic functions $\nu_1$ and $\nu_2$, $\varphi_{\nu_1+\nu_2}(i) = \varphi_{\nu_1}(i) + \varphi_{\nu_2}(i)$ for all $i \in Y$.
\end{enumerate}

In the phylogenetic setting, $\nu(C)$ equals the phylogenetic diversity of $C$ on $T$. 

\begin{definition}[Shapley value (\citet{Haake2008})]
    Let $T$ be a rooted phylogenetic $X$-tree. The \emph{Shapley value} of a leaf $x \in X$ is  
    \begin{align} \label{eq:sv}
        SV_T(x) = \frac{1}{|X|!}  \sum\limits_{C \subseteq X: x \in C} (|C|-1)! (|X|-|C|)! (PD_T(C) - PD_T(C \setminus \{x\})).
    \end{align}
\end{definition}

The computation of the Shapley value for a species $x$ is, at first glance, considerably more involved than that of the fair proportion or equal splits index, as it requires averaging over all subsets $C \subseteq X$ with $x \in C$. However, a striking result first established by \citet{Fuchs2015} shows that, for rooted phylogenetic trees, the Shapley value coincides exactly with the fair proportion index.

\begin{theorem}[\citet{Fuchs2015}] \label{thm:fp-sv}
On rooted phylogenetic $X$-trees, the fair proportion index and the Shapley value coincide. That is, for all $x \in X$,
\[FP_T(x) = SV_T(x).\]
\end{theorem}

A common feature of the PD indices discussed so far (ES, FP, and SV) is their strong dependence on the underlying phylogenetic $X$-tree and its edge lengths. In particular, these indices can be highly sensitive to changes in the tree. For instance, while FP seems robust to random loss of species (\citet{Weedop2019,Perrault2017}), it has recently been shown that the rankings the index produces can be completely reversed when certain species are removed from the tree (see~\citet{Fischer2023,Manson2024} for details). Because removal (extinction) is generally not phylogenetically random, this limitation helps motivate the development of approaches that explicitly incorporate extinction risk into the PD framework. We turn to such extensions in the next section.

\subsection{PD indices incorporating extinction risk} \label{sec:edge2}
While simple phylogenetic diversity indices, such as the fair proportion and equal splits indices, apportion existing phylogenetic diversity among taxa and provide a straightforward basis for prioritization, they do not account for differences in extinction risk among species. Extinction-aware indices extend this framework by incorporating the probability of lineage loss, thereby reflecting the expected contribution of each species to future PD. Under this perspective, a species whose close relatives face high extinction risk carries greater responsibility for preserving shared evolutionary history than a species whose relatives are relatively secure. This concept has been formalized in metrics such as  \emph{conservation potential} (\citet{Weitzman1993}), \emph{heightened evolutionary distinctiveness} (HED), and \emph{heightened evolutionary distinctiveness and globally endangered} (HEDGE) (\citet{Steel2007}), and is further developed in the updated EDGE2 framework through the \emph{ED2} and \emph{EDGE2} metrics (\citet{Gumbs2023}).

To formalize these ideas, we introduce notation that underlies several extinction-aware indices. Let $T$ be a phylogenetic $X$-tree, and suppose that each taxon $x \in X$ has an associated extinction probability $\varepsilon_x$.
We first consider the heightened evolutionary distinctiveness (HED) index.

\paragraph*{Heightened evolutionary distinctiveness index.}
For a taxon $x \in X$ and a subset $Y \subseteq X \setminus \{x\}$, let
\begin{align} \label{eq:pd-complementarity}
    \Delta_{PD}(Y, x) = PD_T(Y \cup \{x\}) - PD_T(Y).
\end{align}
The quantity $\Delta_{PD}(Y, x)$ measures the contribution of $x$ to the phylogenetic diversity of the subtree obtained from $T$ after retaining only the species in $Y$ (for example, following an extinction event). Equivalently, it is the marginal increase in phylogenetic diversity when $x$ is added to $Y$. 

In the special case where $Y=X \setminus \{x\}$, the quantity $\Delta_{PD}(Y,x)$ is also known as \emph{PD complementarity} (\citet{Faith1994}) or the \emph{pendant edge} score (\citet{Altschul1990}). In this setting, it measures the decrease in phylogenetic diversity resulting from the removal of $x$ from the tree, and thus provides a simple way to quantify the contribution of a species to overall PD.

Now let $\mathcal{Y}_x$ be the random subset of species in $X \setminus \{x\}$ that survive a stochastic extinction process, where each species $y \in X \setminus \{x\}$ persists independently with probability $1 - \varepsilon_y$. This model is commonly referred to as the \emph{(generalized) field-of-bullets model} of extinction. 

\begin{definition}[Heightened evolutionary distinctiveness (\citet{Steel2007})]
The heightened evolutionary distinctiveness (HED) of a taxon $x \in X$ is
\begin{align} \label{eq:hed}
    HED_T(x) = \mathbb{E}\big[\Delta_{PD}(\mathcal{Y}_x, x)\big] = \sum\limits_{Y \subseteq X \setminus \{x\}} \mathbb{P}(\mathcal{Y}_x = Y) \Delta_{PD}(Y,x).
\end{align}
\end{definition}
Thus, $HED_T(x)$ quantifies the expected contribution of $x$ to future phylogenetic diversity under extinction risk.

Notice that if all species in $X \setminus \{x\}$ were guaranteed to survive, then $HED_T(x)$ would reduce to the length of the pendant edge incident to leaf $x$. However, under stochastic extinction, $HED_T(x)$ is typically larger than this value, as the loss of close relatives increases the expected marginal contribution of $x$. This \enquote{inflation} motivates the term \emph{heightened} evolutionary distinctiveness (\citet{Steel2007}).

Although the definition of $HED_T(x)$ involves a summation over all subsets of $X \setminus \{x\}$, \citet{Steel2007} showed that it can be computed efficiently on a rooted phylogenetic $X$-tree. In particular,
\begin{align} \label{eq:hed:computation}
   HED_T(x) = \sum_{r=1}^{k_x} \ell(e_r) \cdot \left( \prod_{j \in C(e_r) \setminus \{x\}} \varepsilon_j \right), 
\end{align}
where $e_1, \ldots, e_{k_x}$ are the edges on the path from the root of $T$ to the leaf $x$, and $C(e_r)$ denotes the set of leaves that descend from edge $e_r$. Throughout, we adopt the convention that $\prod_{j \in \emptyset} \varepsilon_j = 1$. Again referring to the tree $T$ in Fig.~\ref{fig:tree} we have, for example, 
\[ HED_T(A) = \ell_{11} \cdot \varepsilon_B \cdot \varepsilon_C + \ell_8 \cdot \varepsilon_B + \ell_1.\]

If extinction does not only occur independently among species, but all species also share a common extinction probability $\varepsilon$, a model known as the \emph{field-of-bullets} (FOB) model, an interesting connection between the HED index and the fair proportion index arises. In particular, \citet{Steel2025} showed that under the FOB model, and assuming a uniform prior $\varepsilon \sim \mathrm{Unif}[0,1]$, we have
\[
FP_T(x) = \mathbb{E}\big[HED_T^\varepsilon(x)\big], \quad \text{for all } x \in X,
\]
where the expectation is taken with respect to the distribution of $\varepsilon$.

\paragraph*{Heightened evolutionary distinctiveness and globally endangered score.}
A closer look at Eq.~\eqref{eq:hed:computation} reveals a subtle but important point: although the HED index of species $x$ accounts for the extinction risk of all other species, it does not explicitly incorporate the extinction risk of $x$ itself, which may initially seem counterintuitive. However, \citet{Steel2007} showed that the HED index of $x$ can be written as the sum of two components, each of which does account for the extinction risk of $x$. Collectively, these components are known as the \emph{heightened evolutionary distinctiveness and globally endangered} (HEDGE) scores.

To make this more precise, let $I_x$ be the random variable that takes the value $x$ if $x$ survives at the future time under consideration, and the empty set otherwise. Define
\begin{align}\label{eq:hedge1}
HED_T'(x) = \mathbb{E}\big[ PD_T(\mathcal{Y}_x \cup \{x\}) - PD_T(\mathcal{Y}_x \cup I_x) \big],
\end{align}
where, as before, $\mathcal{Y}_x$ denotes the random subset of species in $X \setminus \{x\}$ that survive. In other words, $HED_T'(x)$ represents the increase in the expected PD score obtained by conditioning on the survival of species $x$. This quantity is also referred to in the literature as the \emph{(expected) PD complementarity} (\citet{Faith2008}).

Similarly, define 
\begin{align}\label{eq:hedge2}
    HED_T''(x) = \mathbb{E}\big[ PD_T(\mathcal{Y}_x \cup I_x) - PD_T(\mathcal{Y}_x) \big].
\end{align}
In words, $HED_T''(x)$ represents the decrease in expected phylogenetic diversity when we condition on the extinction of species $x$.

\citet{Steel2007} established the following connection between the HEDGE scores $HED_T'(x)$ and $HED_T''(x)$ of species $x$ and its HED index $HED_T(x)$:
\begin{theorem}[\citet{Steel2007}] \label{thm:hed-hedge} \leavevmode
    \begin{enumerate}
        \item $HED_T'(x) = \varepsilon_x \cdot HED_T(x)$,
        \item $HED_T''(x) = (1 - \varepsilon_x) \cdot HED_T(x)$, and
        \item $HED_T'(x) + HED_T''(x) = HED_T(x)$.
    \end{enumerate}
\end{theorem}
This result shows that the HED index naturally decomposes into two complementary contributions, corresponding to the survival and extinction scenarios for species $x$.

Finally, we note that each of $HED_T(x), HED_T'(x)$, and $HED_T(x)''$ is defined as the expectation of a PD-related quantity. Recent work (\citet{Steel2026}) has gone further by studying the corresponding variances of these measures under the generalized field-of-bullets model, establishing relationships similar in spirit to those presented in Theorem~\ref{thm:hed-hedge}.

\paragraph*{EDGE2 protocol.}
While the original EDGE framework dates back to 2007, an interdisciplinary workshop in 2017 initiated a major revision, resulting in the EDGE2 protocol. Key advances include improved methods for handling uncertainty and for incorporating the extinction risk of closely related species (\citet{Gumbs2023}). 

Mathematically, the EDGE2 metric is equivalent to the HEDGE score discussed above. It is framed as a product of two terms resembling the original EDGE score: an evolutionary distinctiveness (ED) component (ED2, the irreplaceability of a species) and an extinction risk component (GE2, the vulnerability of a species). Thus, the EDGE2 score of species $x$ is defined as 
\[ EDGE2(x) = ED2(x) \times GE2(x).\]
According to \citet{Gumbs2023}, $GE2(x)$ is interpreted as a weighting of global endangerment, scaled between 0 and 1, that reflects the relative extinction risk of species $x$. For standardization, the authors set $GE2(x) = \varepsilon_x$, the probability that species $x$ goes extinct by a specified future time. However, they emphasize that this choice is not restrictive, and that alternative measures of extinction risk can be incorporated within the same framework. Moreover, they note that there is currently limited consensus on how best to quantify such extinction probabilities, and propose an approach for incorporating this uncertainty (see \citet{Gumbs2023} for details).

For the ED2 component, the HED index of species $x$ is used; that is, $ED2(x) = HED_T(x)$. In \citet{Gumbs2023}, however, this quantity is expressed by decomposing the sum in Eq.~\eqref{eq:hed:computation} into the contribution of the pendant edge incident to $x$ (which cannot be captured by any other species) and the contributions of the internal edges along the path to the root. Nonetheless, under the choice of $GE2(x) = \varepsilon_x$, we indeed recover the HEDGE score $HED_T'(x)$, since
\begin{align*}
   EDGE2(x) = ED2(x) \times GE2(x) = HED_T(x) \cdot  \varepsilon_x = HED_T'(x),
\end{align*}
where the final equality follows from Theorem~\ref{thm:hed-hedge}.

Finally, notice that the EDGE framework has aimed to include all described species, using rules or imputation methods to handle missing or uncertain phylogenetic information. In EDGE2, this uncertainty is explicitly incorporated by computing ED2 scores across a large distribution of imputed phylogenetic trees, yielding a posterior distribution of scores that allows the robustness of conservation priorities to be assessed (for details, see~\citet{Gumbs2023}).

In summary, the EDGE2 framework extends the original EDGE score by explicitly accounting for the extinction risk of closely related species, while also incorporating uncertainty in both phylogenetic structure and extinction probabilities.

\section{Abundance-weighted measures}\label{sec:abundance}
The PD indices discussed in Sections~\ref{sec:simple} and~\ref{sec:edge2} (such as FP, ES, and HEDGE) assign scores to individual species, reflecting their (expected) contributions to overall phylogenetic diversity. By contrast, abundance-weighted metrics are defined at the community level and quantify the diversity of an assemblage by combining phylogenetic relationships with species’ relative abundances.

A variety of such measures have been proposed (see, e.g., \citet{Leinster2012,Chao2014,Chao2016} for an overview). Here, we focus on two fundamental approaches: Rao's quadratic entropy (\citet{Rao1982}) and phylogenetic Hill numbers (\citet{Chao2010}), which together capture some of the main principles underlying abundance-weighted phylogenetic diversity.

\paragraph*{Rao's quadratic entropy.}
Let $p_i$ denote the relative abundance (e.g., relative number of individuals) of species $i$ in a community of $S$ species, with $p_i \geq 0$ and $\sum_{i=1}^S p_i = 1$. Let $\boldsymbol{p} = (p_1, p_2, \ldots, p_S)^t$ be the corresponding vector of relative abundances. Further, let $\boldsymbol{D} = (d_{ij})_{1 \leq i,j \leq S}$ be a symmetric matrix of pairwise dissimilarities, where $d_{ij} \geq 0$, $d_{ij} = d_{ji}$, and $d_{ii} = 0$ for all $i,j$. In a phylogenetic context, $d_{ij}$ usually represents the patristic distance between species $i$ and $j$ (typically measured as the sum of edge lengths connecting them on a phylogenetic tree), while in a functional context it may represent trait-based dissimilarity.

The \emph{Rao quadratic entropy} (\citet{Rao1982}) of the community is defined as
\begin{align}\label{eq:rao}
    Q(\boldsymbol{q}, \boldsymbol{D}) = \sum_{i=1}^S \sum_{j=1}^S p_i p_j d_{ij}.
\end{align}

Rao's quadratic entropy can be interpreted as the expected dissimilarity between two individuals randomly drawn from the community (with replacement). Equivalently, it is the abundance-weighted average dissimilarity between all pairs of species. 

From Eq.~\eqref{eq:rao}, it is easily seen that Rao’s quadratic entropy is non-negative and increases with both phylogenetic (or functional) dissimilarity among species and the evenness of their relative abundances. 
For further details on the mathematical properties of Rao's quadratic entropy, see, e.g., \citet{Pavoine2005,Pavoine2026}, and the references therein.

As we shall see next, Rao’s quadratic entropy corresponds to a particular case within the broader class of phylogenetic Hill numbers, which provide a unified framework for abundance-weighted diversity by tuning the sensitivity to species’ relative abundances.

\paragraph*{Phylogenetic Hill numbers.}
While Rao's quadratic entropy provides a single abundance-weighted measure based on pairwise dissimilarities, it is often desirable to consider a family of diversity measures that vary in their sensitivity to species’ relative abundances.

Such a family is given by the \emph{phylogenetic Hill numbers} (\citet{Chao2010}), which extend the standard Hill numbers (effective number of species) to a phylogenetic setting. 

For $q \geq 0$ with $q \neq 1$, the standard Hill number of order $q$ for a community of $S$ species is defined as
\begin{align}\label{eq:hill-standard}
    D^q = \left( \sum_{i=1}^S p_i^q \right)^{1/(1-q)},
\end{align}
where $p_i$ denotes the relative abundance of species $i$.
The case $q=1$ is defined by continuity, and the limit as $q \to 1$ yields
\begin{align*}
    D^1 = \lim_{q \rightarrow 1} D^q = \exp \left(- \sum_{i=1}^S p_i \log p_i \right),
\end{align*}
which corresponds to the exponential of Shannon entropy.

In all cases, the parameter $q$ controls the sensitivity of the measure to species abundances, with larger values of $q$ placing increasing emphasis on more abundant species.

Phylogenetic Hill numbers generalize this definition by incorporating the shared evolutionary history among species through an underlying (often ultrametric) phylogenetic tree. Consider a time interval $[-t,0]$, where time $0$ denotes the present and $t$ is the time before present. Following \citet{Chao2010}, the \emph{phylogenetic Hill number of order $q$ over the interval $[-t,0]$}, also referred to as the \emph{mean phylogenetic diversity of order $q$}, is defined as
\begin{align}
\overline{D}^q(t)
= \left( \sum_{i \in B_t} \frac{\ell_i}{t} \, a_i^q \right)^{1/(1-q)},
\qquad q \geq 0,\; q \neq 1,
\end{align}
and
\begin{align}
\overline{D}^1(t)
= \lim_{q \to 1} \overline{D}^q(t)
= \exp\left( - \sum_{i \in B_t} \frac{\ell_i}{t} \, a_i \log a_i \right).
\end{align}

Here, $B_t$ denotes the set of edges in the time interval $[-t,0]$, $\ell_i$ is the length of edge $i$, and $a_i$ is the total relative abundance descended from edge $i$, defined as the sum of the relative abundances of all species subtended by that edge. For further details and worked examples, see \citet{Chao2010}.

Several important special cases link phylogenetic Hill numbers to the measures discussed above. In particular, when $t$ is chosen as the age of the root, we obtain (see, e.g., \citet{Chao2016} for details):
\begin{itemize}
    \item For $q=0$, all lineages are weighted equally, and $\overline{D}^0(t)$ reduces (up to a normalization by $t$) to Faith's phylogenetic diversity (PD).
    \item For $q=2$, the measure corresponds to a quadratic entropy-type index and is closely related to Rao's quadratic entropy, placing greater emphasis on abundant lineages.
\end{itemize}

\paragraph*{Originality of a species.}
Both Rao's quadratic entropy and phylogenetic Hill numbers quantify the diversity of a community or a set of species rather than the contribution of individual taxa. As with PD, however, species-specific indices may be derived from these community-level measures.

For Rao's quadratic entropy, this idea was pioneered by \citet{Pavoine2005QE}, who introduced the concept of the \enquote{originality of a species within a set}. Their \emph{quadratic entropy (QE)-based originality index} defines the originality of a species as its optimal relative abundance in the abundance distribution that maximizes Rao's quadratic entropy on an ultrametric phylogenetic tree. Species assigned higher optimal abundances are interpreted as making larger contributions to the maximum attainable quadratic entropy and are therefore considered more original. Unlike indices like fair proportion and equal splits, which allocate edge lengths directly to species, the QE-based originality index derives species scores from an optimization problem on Rao's quadratic entropy.

For the phylogenetic Hill numbers, on the other hand, we are not aware of a generally accepted species-level analogue. Extending this framework to species-specific indices may be an interesting direction for future research.

\bigskip
All measures of biodiversity and their associated indices discussed so far assume that evolutionary relationships are adequately represented by a phylogenetic tree and that diversity can be expressed through additive edge lengths
However, this tree-based representation may be overly restrictive in situations where evolutionary histories involve reticulation processes such as hybridization, recombination, lateral gene transfer, or even gene flow among populations. In such cases, phylogenetic relationships are more appropriately represented by networks rather than trees, motivating extensions of phylogenetic diversity measures to phylogenetic networks.

\section{Beyond trees: PD and PD indices on phylogenetic networks} \label{sec:networks}

Evolutionary relationships among species have traditionally been represented using phylogenetic trees. However, it is now widely recognized that evolution is not always strictly tree-like; many biological systems experience events in which genetic material is transferred \emph{horizontally} between taxa, rather than \emph{vertically} from ancestor to descendant. Such events include hybridization, lateral gene transfer, and recombination.

As a result, phylogenetic networks are increasingly used to capture these more complex patterns of evolution. Correspondingly, extensions of phylogenetic diversity (PD) and PD indices to networks have been proposed. Although research in this area remains active and ongoing, in this section, we review some recent developments, discussing both implicit and explicit phylogenetic networks.

\subsection{Implicit phylogenetic networks}
Implicit, or abstract, phylogenetic networks are typically undirected and are used to represent the extent to which data deviate from a strictly tree-like structure. In particular, they are well suited for visualizing conflicting evolutionary signals.

Among the most widely used implicit networks are \emph{split networks}, which can be constructed from pairwise distances between taxa using the well-known \emph{Neighbor-Net} algorithm (\citet{Bryant2003}). 
A split corresponds to a bipartition of the taxa into two groups, representing a possible evolutionary separation. In a purely tree-like scenario, these splits are all mutually compatible and can be represented by a single tree. However, when the data contain conflicting signals, different splits may be incompatible with one another.

\begin{figure}[htbp]
    \centering
    \includegraphics[width=0.8\linewidth]{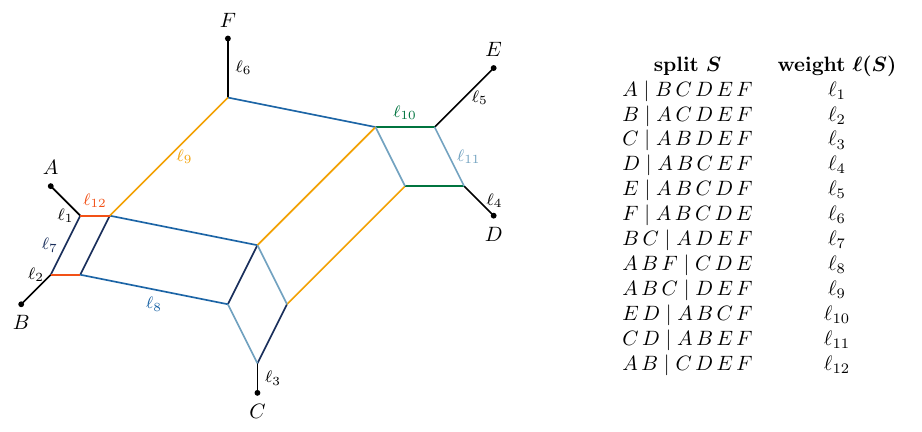}
    \caption{A split network visualizing a collection of weighted splits on the set $X = \{A, \ldots, F\}$. For example, the parallel edges in light orange represent the bipartition separating taxa $A,B,C$ from $D,E,F$, denoted by the split $A \, B \, C \mid D \, E\, F$, with associated weight $\ell_9$ (set braces are omitted for readability).}
    \label{fig:splitnet}
\end{figure}

A split network provides a way to visualize all such splits simultaneously. For example, if one subset of the data suggests grouping taxa $A$ and $B$ together, while another supports grouping $B$ with $C$, a tree can only display one of these relationships. In contrast, a split network represents both signals, typically through parallel edges or box-like structures that highlight the ambiguity. An example is shown in Figure~\ref{fig:splitnet}.

Split networks are widely applied in biological research and can, for instance, be obtained through the popular software \texttt{SplitsTree} (\citet{Huson2005,Huson2024}).

Both PD itself and several of the PD indices introduced earlier have been extended to the setting of weighted split networks, or more generally, weighted split systems on $X$. Formally, a \emph{split} $S$ of $X$ is a bipartition of $X$ into two non-empty subsets $X_1$ and $X_2$, that is, $X_1 \cup X_2 = X$ and $X_1 \cap X_2 = \emptyset$. Such a split is denoted by the unordered pair $X_1 \mid X_2 = X_2 \mid X_1$. A \emph{split system} $\mathcal{S}$ on $X$ is a non-empty collection of splits of $X$. We further assume that each split $S \in \mathcal{S}$ is assigned a non-negative weight $\ell(S)$, leading to a weighted split system.

\paragraph*{Phylogenetic diversity for split systems.}
Let $\mathcal{S}$ be a weighted split system on $X$, and let $Y \subseteq X$. Then, the phylogenetic diversity induced by $\mathcal{S}$ on $Y$ is defined as
\begin{align} \label{eq:split-pd}
    PD_\mathcal{S}(Y) &= \sum\limits_{\substack{X_1 \mid X_2 \in \mathcal{S}\\ X_1 \cap Y \neq \emptyset, X_2 \cap Y \neq \emptyset}} \ell(X_1 \mid X_2),
\end{align}
that is, we sum the weights of all splits that separate at least one element of $Y$ on each side. 

As an example, consider the split network in Figure~\ref{fig:splitnet}, and let $Y = \{A,B,C\}$. The splits that separate the taxa in $Y$ are $\{A\} \mid X \setminus \{A\}$, $\{B\} \mid X \setminus \{B\}$, $\{C\} \mid X \setminus \{C\}$, $\{B,C\} \mid X \setminus \{B,C\}$, $\{A,B,F\} \mid \{C,D,E\}$, and $\{A,B\} \mid X \setminus \{A,B\}$. Hence, $PD_\mathcal{S}(Y) = \ell_1 + \ell_2 + \ell_3 + \ell_7 + \ell_8 + \ell_{12}$.

A natural optimization problem, analogous to the tree setting, is to select a subset of species $Y \subseteq X$ of cardinality $k$ that maximizes phylogenetic diversity. While this problem is NP-hard for general split systems (\citet{Spillner2008}), it becomes tractable in special cases. In particular, if the split system is circular (as is the case for those produced by the \emph{Neighbor-Net} algorithm) or, more generally, affine, the problem can be solved efficiently (\citet{Spillner2008,Minh2009}). 

However, it is important to keep in mind that empirical data do not necessarily give rise to circular or affine split systems, and enforcing such structure may distort phylogenetic diversity calculations (see \citet{Abhari2024} for a more detailed discussion).

\paragraph*{Phylogenetic diversity indices for split systems.}
With a definition of PD for split systems in hand (Eq.~\eqref{eq:split-pd}), any PD-based index defined directly in terms of PD on subsets of taxa, such as the Shapley value (Eq.~\eqref{eq:sv}) or the HED index (Eq.~\eqref{eq:hed}), extends naturally to this setting. In both cases, alternative methods have been developed that sum over all splits rather than over all subsets of taxa, leading to potentially more efficient implementations.

Let $\mathcal{S}$ be a collection of splits on $X$, and let $x \in X$. The Shapley value of $x$ with respect to $\mathcal{S}$ is given by (\citet{Haake2008,Volkmann2014})
\begin{align*}
    SV_\mathcal{S}(x) = \sum\limits_{X_1 \mid X_2 \in \mathcal{S}: x \in X_1} \frac{\vert X_2 \vert}{\vert X \vert \cdot \vert X_1 \vert} \ell(X_1 \mid X_2),
\end{align*}
where $\ell(X_1 \mid X_2)$ denotes the weight of split $X_1 \mid X_2$. Both PD for split systems and the corresponding Shapley value can be computed using \texttt{SplitsTree} (\cite{Huson2024}).

As an example, consider taxon $A$ in the split network shown in Figure~\ref{fig:splitnet}. Then,
\begin{align*}
    SV_\mathcal{S}(A) &= \frac{5}{6 \cdot 1} \ell_1 + \frac{4}{6 \cdot 2} \ell_{12}+ \frac{3}{6 \cdot 3} \left( \ell_8 + \ell_9 \right) + \frac{2}{6 \cdot 4} \left( \ell_7 + \ell_{10} + \ell_{11} \right) \\
     &\qquad + \frac{1}{6 \cdot 5} \left( \ell_2 + \ell_3 + \ell_4 + \ell_5 + \ell_6 \right)\\ 
     &= \frac{5}{6} \ell_1 + \frac{1}{3} \ell_{12} + \frac{1}{6} \left( \ell_8 + \ell_9 \right) + \frac{1}{12} \left( \ell_7 + \ell_{10} + \ell_{11} \right) \\
     &\qquad + \frac{1}{30} \left( \ell_2 + \ell_3 + \ell_4 + \ell_5 + \ell_6 \right).
\end{align*}

Similarly, the HED index of $x$ with respect to $\mathcal{S}$ is given by (\citet{Volkmann2014})
\begin{align*}
    HED_{\mathcal{S}}(x) = \sum\limits_{X_1 \mid X_2 \in \mathcal{S}: x \in X_1} \left( \prod\limits_{y \in X_1 \setminus \{x\}} \varepsilon_y \right) \cdot \left(1 - \prod\limits_{y \in X_2} \varepsilon_y \right) \cdot \ell(X_1 \mid X_2),
\end{align*}
where, as in Section~\ref{sec:edge2}, $\varepsilon_x$ denotes the probability that species $x$ goes extinct by a specified future time, and we use the convention that $\prod_{j \in \emptyset} \varepsilon_j=1$.

Referring again to Figure~\ref{fig:splitnet}, we have, for example, 
\begin{align*}
    HED_{\mathcal{S}}(A) &= 1 \cdot (1 - \varepsilon_B \varepsilon_C \varepsilon_D \varepsilon_E \varepsilon_F) \cdot \ell_1
    + \varepsilon_C \varepsilon_D \varepsilon_E \varepsilon_F \cdot (1 - \varepsilon_B) \cdot \ell_2\\
    &\quad + \varepsilon_B \varepsilon_D \varepsilon_E \varepsilon_F \cdot (1 - \varepsilon_C) \cdot \ell_3
    + \varepsilon_B \varepsilon_C \varepsilon_E \varepsilon_F \cdot (1 - \varepsilon_D) \cdot \ell_4 \\
    &\quad + \varepsilon_B \varepsilon_C \varepsilon_D \varepsilon_F \cdot (1 - \varepsilon_E) \cdot \ell_5 
    + \varepsilon_B \varepsilon_C \varepsilon_D \varepsilon_E \cdot (1 - \varepsilon_F) \cdot \ell_6 \\
    &\quad + \varepsilon_D \varepsilon_E \varepsilon_F \cdot (1 - \varepsilon_B \varepsilon_C) \cdot \ell_7
    + \varepsilon_B \varepsilon_F \cdot (1 - \varepsilon_C \varepsilon_D \varepsilon_E) \cdot \ell_8\\
    &\quad + \varepsilon_B \varepsilon_C \cdot (1 - \varepsilon_D \varepsilon_E \varepsilon_F) \cdot \ell_9 
    + \varepsilon_B \varepsilon_C \varepsilon_F \cdot (1 - \varepsilon_D \varepsilon_E) \cdot \ell_{10}\\
    &\quad + \varepsilon_B \varepsilon_E \varepsilon_F \cdot (1 - \varepsilon_C \varepsilon_D) \cdot \ell_{11}
    + \varepsilon_B \cdot (1 - \varepsilon_C \varepsilon_D \varepsilon_E \varepsilon_F) \cdot \ell_{12}.
    \end{align*}

Finally, we note that recent work by \citet{Moulton2024} extends indices such as the fair proportion index and the equal splits index to weighted split systems, and also introduces a general theoretical framework for deriving PD-based indices from collections of splits or clusters. The latter arise naturally in the context of explicit phylogenetic networks, which we consider in the next subsection.

\subsection{Explicit phylogenetic networks}
Explicit (or rooted) phylogenetic networks are directed graphs that aim to represent how a group of species evolved from a common ancestor through a combination of tree-like processes and reticulation events such as hybridization or lateral gene transfer. A wide range of subclasses of rooted phylogenetic networks has been introduced and studied in the literature, each characterized by specific structural constraints motivated by biological realism, mathematical tractability, or a combination of both (\citet{Kong2022,Huson2011,Steel2016}).

Formally, a \emph{rooted phylogenetic network} $N$ on $X$ is a rooted directed acyclic graph with no parallel edges satisfying the following properties:
\begin{enumerate}
\item The (unique) root has in-degree zero and out-degree at least two;
\item A vertex with out-degree zero has in-degree one, and the set of all such vertices is $X$ (these are the leaves);
\item Every other vertex is either a \emph{tree vertex}, having in-degree one and out-degree at least two, or a \emph{reticulation}, having in-degree at least two and out-degree one.
\end{enumerate}

\begin{figure}[htbp]
    \centering
    \includegraphics[width=.5\linewidth]{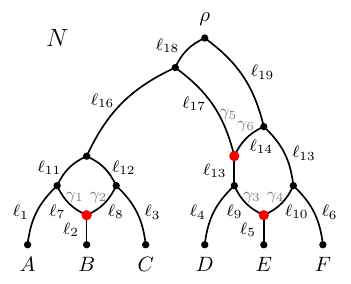}
    \caption{A rooted binary phylogenetic network $N$ on $X = \{A, \ldots, F\}$. All edges are directed downward. Leaves represent observed species, and internal nodes are either reticulations (red vertices) or tree nodes (black vertices). Edges length $\ell_1, \ldots, \ell_{19}$ indicate the amount of evolutionary change or time separating lineages, while inheritance probabilities $\gamma_1, \ldots, \gamma_6$ represent the proportion of genetic material (or features) that a reticulation node inherits from its two parents.}
    \label{fig:net}
\end{figure}

If \enquote{at least} is replaced by \enquote{exactly} in the above degree conditions, the network is said to be \emph{binary}. An example of a rooted binary phylogenetic network is shown in Figure~\ref{fig:net}.

We refer to edges directed into a reticulation as \emph{reticulation edges} and to all other edges as \emph{tree edges}. Furthermore, throughout this section, we assume that all edges of $N$ are assigned non-negative real-valued lengths. That is, if $E$ denotes the edge set of $N$, then there is a mapping $\ell$: $E \rightarrow \mathbb{R}_{\geq 0}$ that assigns to each edge $e \in E$ a weight $\ell(e)$. 

As in the case of phylogenetic trees, the interpretation of these edge lengths is context-dependent and may represent, for example, the expected number of substitutions per site, coalescent units, or elapsed (calendar) time. 

Additionally, if $N$ is binary and $v$ is a reticulation with in-coming edges $e_1 = (u_1, v)$ and $e_2 = (u_2,v)$, we may associate inheritance probabilities $\gamma_{e_1}$ and $\gamma_{e_2}$ with $e_1$ and $e_2$, respectively. These values represent the proportion of genetic material (or features) that vertex $v$ inherits from its parents $u_1$ and $u_2$. In some contexts, these probabilities are required to sum to one. Extensions to reticulations with higher in-degree are straightforward.

\paragraph*{Notions of phylogenetic diversity for rooted phylogenetic networks.}
Recall that the PD of a subset $Y \subseteq X$ of taxa on a rooted phylogenetic $X$-tree is defined as the sum of edge lengths in the minimal subtree connecting the taxa in $Y$ to the root. Several generalizations of PD to rooted phylogenetic networks have been proposed in the literature. These include:
\begin{itemize}
    \item \emph{Phylogenetic (sub)net diversity} (\citet{Wicke2018}) or \emph{AllPaths-PD} (\citet{Bordewich2022}), which for a subset $Y \subseteq X$ of taxa in a rooted phylogenetic network $N$ on $X$ is defined as the sum of the lengths of all edges in $N$ that lie on a directed path from the root of $N$ to a leaf in $Y$, that is,
    \begin{align*}
        \textup{AllPaths-PD}_N(Y) = \sum\limits_{e \in \text{Anc}(Y)} \ell(e),
    \end{align*}
    where $\text{Anc}(Y)$ denotes the set of edges of $N$ that are ancestral to at least one taxon in $Y$, i.e., that lie on a directed path from the root to some leaf in $Y$.

    \item \emph{Network-PD} (\citet{Bordewich2022}), a measure that incorporates the fact that, at a reticulation event, a taxon may inherit only a fraction of its genetic material (or features) from each parent (assuming that reticulation edges are equipped with inheritance probabilities). For a subset \(Y\) of the leaves of \(N\), define, for each edge \(e = (u,v)\), the quantity \(\gamma(Y,e)\) to be the proportion of the features of \(v\) that are present in the taxa in \(Y\). Equivalently, \(\gamma(Y,e)\) is the probability that a feature arising on edge \(e\) is inherited by at least one taxon in \(Y\). We then define
    \begin{align*}
    \textup{Network-PD}_N(Y) = \sum_{e \in N} \gamma(Y,e)\,\ell(e).
    \end{align*}
    The values \(\gamma(Y,e)\) can be computed in a bottom-up manner (\citet{Bordewich2022}). Moreover, when all inheritance probabilities are equal to \(1\), Network-PD coincides with AllPaths-PD (\citet{Bordewich2022}).

    \item Measures based on trees embedded in the network, in particular \emph{MaxWeightTree-PD}, \emph{MinWeightTree-PD} (\citet{Bordewich2022}) defined by
    \begin{align*}
    \textup{MaxWeightTree-PD}_N(Y) &= \max_{T \in \mathcal{T}_Y(N)} \sum_{e \in T} \ell(e), \\
    \textup{MinWeightTree-PD}_N(Y) &= \min_{T \in \mathcal{T}_Y(N)} \sum_{e \in T} \ell(e).
    \end{align*}
    Here, $\mathcal{T}_Y(N)$ denotes the set of connecting subtrees for $Y$ in $N$, where a subgraph $T$ of $N$ is called a connecting subtree if it is a rooted directed tree with root $\rho$ and leaf set $Y$. The measure MinWeightTree-PD is also referred to as \emph{phylogenetic net diversity} by \citet{Wicke2018}, where additional notions of \enquote{embedded PD} are discussed. 

    A related notion is \emph{AverageTree-PD} (\citet{vanIersel2025}), defined by 
    \begin{align*}
        \textup{AverageTree-PD}_N(Y) = \sum\limits_{T \in \mathcal{T}_Y(N)} \mathbb{P}(T) \cdot PD_T(Y), 
    \end{align*}
    where $\mathbb{P}(T)$ denotes the probability of the connecting subtree $T$, obtained as the product of the inheritance probabilities associated with the reticulation edges retained in $T$, and $PD_T(Y)$ is the sum of edge lengths of $T$.
\end{itemize}

Analogous to the setting of rooted phylogenetic $X$-trees or split networks on $X$, a natural optimization problem is to compute the maximum PD score over all subsets of taxa of size $k$. In the context of rooted phylogenetic networks, this problem is NP-hard for most of the measures discussed above, including AllPaths-PD, Network-PD, MinWeightTree-PD, and AverageTree-PD (\citet{Bordewich2022}). For the latter two measures, even computing the PD value for a fixed subset $Y \subseteq X$ is NP-hard (\citet{Bordewich2022,vanIersel2025}. On the positive side, the problem is solvable in polynomial time for MaxWeightTree-PD (\citet{Bordewich2022}). Furthermore, for certain variants of PD and restricted classes of networks, additional positive results have been obtained, alongside a growing body of work on parameterized algorithms for these problems (see, e.g., \citet{Bordewich2022,Jones2023,vanIersel2025,vanIersel2025b,Coronado2024} and references therein).

Furthermore, we remark that several of the above measures have recently been implemented in \texttt{PaNDA} (Phylogenetic Network
Diversity Algorithms), a Python software package and interactive graphical user-interface for exploring, visualizing and maximizing diversity in phylogenetic networks (\citet{Holtgrefe2025}).

\begin{remark}
A relatively recent class of phylogenetic networks are \emph{semi-directed} networks, which lie between rooted (directed) and unrooted (undirected) phylogenetic networks in that they contain both directed and undirected edges. Informally, they can be obtained from rooted phylogenetic networks by suppressing the root and undirecting all tree edges, while retaining the direction of reticulation edges. Such networks arise naturally in practice, as several widely used inference methods return semi-directed rather than fully rooted networks. Very recently, AllPaths-PD has been extended to this setting (\citet{Holtgrefe2025}). The authors further show that finding a subset of $k$ taxa with maximum AllPaths-PD remains NP-hard. However, they present a polynomial-time algorithm for networks of bounded level, where the level is a well-known measure of tree-likeness, defined as the maximum number of edges that must be removed from any \emph{blob} (i.e., reticulated or biconnected component) to obtain a tree.
\end{remark}

\paragraph*{Phylogenetic diversity indices for rooted phylogenetic networks.}
As in the case of split systems discussed above, any PD-based index defined directly in terms of PD on subsets of taxa, such as the Shapley value (Eq.~\eqref{eq:sv}) or the HED index (Eq.~\eqref{eq:hed}), can be applied to any of the network-based notions of PD introduced above (see, e.g., \citet{Wicke2018}). However, the computation of these indices may quickly become intractable for networks containing more than a few taxa.

In the case of AllPaths-PD, an interesting connection arises between the Shapley value and a generalized fair proportion index, analogous to Theorem~\ref{thm:fp-sv}. Specifically, \citet{Coronado2018} define the following extension of the fair proportion index. Let $N$ be a rooted phylogenetic network on $X$. The fair proportion index of a leaf $x \in X$ is given by
\begin{align} \label{eq:fp-net}
FP_N(x) = \sum_{e \in \textup{Anc}(\{x\})} \frac{\ell(e)}{n(e)},
\end{align}
where $n(e)$ denotes the number of leaves descended from the edge $e$, and $\textup{Anc}({x})$ is the set of edges that are ancestral to $x$.

As an example, consider the network $N$ shown in Figure~\ref{fig:net}. For instance, we have
\[FP_N(B) = \frac{\ell_2}{1} + \frac{\ell_7}{1} + \frac{\ell_8}{1} + \frac{\ell_{11}}{2} + \frac{\ell_{12}}{2} + \frac{\ell_{16}}{3} + \frac{\ell_{18}}{5}.\]

Now, let 
\begin{align*}
    SV^\textup{AllPaths-PD}_N(x) &= \frac{1}{|X|!}  \sum\limits_{C \subseteq X: x \in C} (|C|-1)! (|X|-|C|)! (\textup{AllPaths-PD}_N(C)\\
    &\qquad- \textup{AllPaths-PD}_N(C \setminus \{x\})).
\end{align*}
denote the Shapley value of $x$ induced by AllPaths-PD on $N$.

\begin{theorem}[\citet{Coronado2018}] \label{thm:fp-sv-net}
For every weighted phylogenetic network $N$ on $X$, the network-based fair proportion index coincides with the Shapley value induced by AllPaths-PD. That is, for all $x \in X$, 
\begin{align*}
FP_N(x) &= SV^\textup{AllPaths-PD}_N(x).
\end{align*}
\end{theorem}

Finally, we remark that the PD indices for general cluster systems introduced by \citet{Moulton2024}, including a generalization of the equal splits index (Eq.~\eqref{eq:es}), are also applicable to rooted phylogenetic networks, as each such network naturally induces a cluster system. Specifically, if $N$ is a rooted phylogenetic network on $X$, then each vertex $v$ of $N$ induces a \emph{(hardwired) cluster}, namely the set of taxa descended from $v$. The collection of all such clusters forms a cluster system. Moreover, weights for these clusters can be derived directly from the edge lengths of the network.

\medskip
In summary, a variety of extensions of phylogenetic diversity and associated indices have been developed for both implicit and explicit phylogenetic networks. These approaches provide different ways of capturing evolutionary relationships in the presence of reticulation or conflicting signal, and often lead to distinct mathematical and computational properties. However, much of the existing work has been driven by mathematical and theoretical computer science considerations. An important direction for future research is therefore to assess the practical performance of these measures and to determine which notions of phylogenetic diversity on networks are most informative and biologically meaningful in empirical applications.

\section{Interpretation of edge lengths} \label{sec:lengths}
In previous sections, we reviewed phylogenetic diversity (PD) and PD indices for both trees and networks. Although these measures differ in their definitions and properties, they all ultimately depend on the edge lengths of a phylogenetic tree or network. Optimizing PD, therefore, amounts to selecting species that collectively capture as much of this total edge length as possible. This raises a fundamental question: what do these edge lengths actually represent, and what, precisely, are we preserving when we aim to conserve PD? 

In what follows, we focus on phylogenetic trees for simplicity, although many of the considerations extend, with appropriate modifications, to phylogenetic networks.

While PD and PD indices are defined in terms of edge lengths, there are different ways of interpreting these measures. In practice, edge lengths may represent different biological or statistical quantities depending on how the phylogenetic tree is inferred. In \emph{chronograms} (time-calibrated trees), they correspond to divergence times, whereas in so-called \emph{phylograms}, they typically represent the expected number of substitutions per site under a specified model of sequence evolution (see, e.g.,~\citet{Felsenstein2004}) for specified sequences. In other settings, particularly under the \emph{multispecies coalescent} (e.g., \citet{Rannala2003,Degnan2005,Degnan2009}), a standard model for studying \emph{incomplete lineage sorting} (a process by which trees inferred for individual genes may differ from the species tree), edge lengths may be expressed in coalescent units, reflecting time to lineage coalescence scaled by effective population size. Importantly, these different representation are not simple rescalings of one another. For example, coalescent units incorporate population-level processes and may differ substantially from absolute time when effective population sizes vary across lineages. More generally, edge lengths are model-dependent estimates, reflecting both the data and the assumptions used in phylogenetic reconstruction.

This ambiguity gives rise to multiple interpretations of what PD measures and, consequently, what is being preserved when PD is maximized. One interpretation is that PD captures the total amount of evolutionary history represented by a set of taxa, particularly when edge lengths are proportional to time. Under this interpretation, maximizing PD corresponds to preserving as much \enquote{evolutionary time} as possible. In its original formulation, however, PD was motivated by a model in which shared ancestry explains shared observed features, allowing PD to be interpreted as predicting the diversity of features represented by a set of species~(\citet{Faith1992}). This interpretation naturally leads to a second viewpoint, in which PD is linked to measured functional or feature diversity. Under a simple model in which new features arise along the edges of a tree and are never lost, the features present in a set of species $S$ (its feature diversity, $FD(S)$) are indeed captured by the phylogenetic diversity of $S$ (\citet{Wicke2021}). However, when feature loss is allowed, recent results show that under both deterministic and stochastic models of feature gain and loss, FD necessarily deviates from PD except in very restricted cases involving trivial tree topologies (see, e.g., \citet{Overwater2023,Rosindell2023}). 

More broadly, the extent to which PD accurately reflects functional diversity at clade-level and local assembly scales has been widely debated in the biological literature (see, e.g., \citet{Devictor2010,Mazel2018,Mazel2019,Owen2019,Tucker2018,Tucker2019}). The debate has two related foci: first, whether features that we deem currently valuable (e.g., those that contribute to ecological function) are expected to diverge over time such that a phylogeny (ultrametric or additive) is a good predictor of their distribution, or whether convergent or parallel evolution for such desirable features is common; and second, how the processes that lead to local community assembly (environmental filtering, species interactions, and biogeographic history) 
mediate the overall phylogenetic signal of functional traits at local, conservation-relevant scales. Recent work in plant communities (e.g., \citet{Hahn2024, Vecera2023}) suggests that these latter processes can lead to decoupling of functional and phylogenetic diversity at local scales, complicating the predictive power of phylogenies. 

A third interpretation, still linked to the idea that PD captures feature diversity, views PD as a proxy for \enquote{option value}: the potential future benefits that biodiversity may provide, particularly currently unknown uses or ecological functions. As \citet{Faith2015} notes, \enquote{Biodiversity conservation maintains ‘option values’---potential unanticipated future benefits, from various specific elements of biodiversity.} From this perspective, preserving lineages that span a large portion of the Tree of Life increases the likelihood of retaining rare or unique genetic, biochemical, or ecological properties precisely because the underlying tree predicts the distribution of unmeasured features. Although appealing, this interpretation is more heuristic and less directly tied to directly measurable quantities.

While these three interpretations of PD outlined above may not be exhaustive, they highlight that PD is a measure whose meaning depends on how edge lengths are understood. This issue is compounded by the fact that estimating edge lengths itself is a challenging statistical problem. While considerable attention has been devoted to inferring tree topology, edge length estimates are often more sensitive to model misspecification, rate heterogeneity among lineages, and uncertainty in calibration or population parameters (see, e.g., \citet{Schwartz2010,Brown2009,Fleming2023,Suvorov2024}). As a result, the values of PD, and the rankings of taxa or sets of taxa based on PD and PD indices, may be substantially affected by uncertainty in both the topology of the underlying phylogeny as well as its edge lengths.

In summary, these considerations suggest that the use of PD and PD indices as a criterion for prioritization relies on implicit assumptions about the interpretation and reliability of edge lengths. Making these assumptions explicit is therefore essential when applying PD in empirical or conservation contexts, particularly when it is used as a proxy for other aspects of biodiversity such as evolutionary time, feature, or functional diversity. We believe this proxy use is almost universal.

\section{Concluding remarks} \label{sec:future}
In this chapter, we have reviewed the development of phylogeny-based conservation metrics stemming from Daniel P. Faith's 1992 study (\citet{Faith1992}), which introduced the concept of phylogenetic diversity (PD). This foundational paper has led to an extensive methodological framework of PD indices, all aimed at conserving the \enquote{Tree (or Network) of Life.}

As this review shows, the field has undergone continuous development. This progress is reflected, for example, in the transition from simple PD indices based solely on phylogenetic trees and their edge lengths to those that incorporate extinction risk and uncertainty, as well as the extension from phylogenetic trees to phylogenetic networks. 

We expect this development to continue, particularly in the analysis and extension of PD indices under more sophisticated models of extinction and in more general settings beyond tree-based PD. In a collaboration with Mike Steel (\citet{Steel2026}), for example, we have recently shown that the expected diversity framework from PD measured on trees and networks can be generalized to a setting where measured features are arbitrarily associated with taxa (see also \citet{Wicke2021}). This extends the \enquote{PD calculus} but makes the metrics less predictive because the distribution of the measured features are not expected to correlate with the distribution of unmeasured features (contra the underlying justification for using PD in the \enquote{option value} context); however, the generalization may open up new avenues for conservation biologists. We have also derived the variance of the general class of indices, which allows for the quantification of uncertainty associated with these scores, useful when comparing species or prioritizing conservation actions under stochastic extinction scenarios.  We expect further development in this direction.

Several other avenues for future research remain. A first example is the systematic incorporation of uncertainty, not only in species’ extinction risks but also in the topology and edge lengths of phylogenies, into PD-based metrics, moving from deterministic scores to approaches that quantify variability. In addition, while a variety of PD generalizations for phylogenetic networks have recently been proposed and analyzed from a theoretical computer science perspective, an important question is to assess their mathematical properties, robustness, and biological interpretability, and to determine which variants are most appropriate in practice. In fact, this challenge is not specific to network-based variants. Even for phylogenetic trees, a wide range of diversity indices and PD generalizations have been proposed (of which we have presented only a selection), and further work is needed to systematically compare their properties and biological relevance in different contexts (see also \citet{Pavoine2026b} for a related discussion). Further work is also needed in integrating more realistic models of extinction, including dependencies among species and environmental factors, as well as in developing scalable methods for computing PD associated quantities, capable of handling increasingly large and complex phylogenetic data sets.

\bibliographystyle{plainnat}
\bibliography{References_WickeMooers}

\end{document}